\documentclass{iaus}
\usepackage{graphicx}
\usepackage{natbib}
\usepackage{epsf}

\def\araa{{\em ARAA}}
\def\aj{{\em AJ}}
\def\apj{{\em ApJ}}
\def\apjl{{\em ApJ}}
\def\apjs{{\em ApJS}}
\def\aap{{\em A\&A}}

\def\mnras{{\em MNRAS}}

\def\pasp{{\em PASP}}

\title[Gravoturbulent Star Formation] %% give here short title %%
{Gravoturbulent Star Formation: Effects of the Equation of State on
  Stellar Masses}
\author[Klessen, Spaans, \& Jappsen]   %% give here short author list %%
{Ralf S.\ Klessen$^1$, Marco Spaans$^2$, Anne-Katharina Jappsen$^1$}

\affiliation{$^1$ Astrophysikalisches Institut Potsdam, An der Sternwarte 16, 14482
  Potsdam, Germany \break email: rklessen@aip.de, akjappsen@aip.de\\[\affilskip]
$^2$ Kapteyn Astronomical Institute, P.O. Box 800, 9700 AV Groningen, The Netherlands  \break email: spaans@astro.rug.nl}

\pubyear{2005}
\volume{xxx}  %% insert here IAU Symposium No.
\pagerange{119--126}
\date{?? and in revised form ??}
\jname{Proceedings Title IAU Symposium}
\editors{A.C. Editor, B.D. Editor \& C.E. Editor, eds.}
\begin{document}

\maketitle

\begin{abstract}
  Stars form by gravoturbulent fragmentation of interstellar gas
  clouds.  The supersonic turbulence ubiquitously observed in Galactic
  molecular gas generates strong density fluctuations with gravity
  taking over in the densest and most massive regions.  Collapse sets
  in to build up stars and star clusters.
  
  Turbulence plays a dual role. On global scales it provides support,
  while at the same time it can promote local collapse.  Stellar birth
  is thus intimately linked to the dynamic behavior of parental gas
  clouds, which governs when and where protostellar cores form, and how
  they contract and grow in mass via accretion from the surrounding
  cloud material to build up stars.  The equation of state plays a
  pivotal role in the fragmentation process.  Under typical cloud
  conditions, massive stars form as part of dense clusters following
  the ``normal'' mass function observed, e.g.\ in the solar
  neighborhood.  However, for gas with an effective polytropic index
  greater than unity star formation becomes biased towards isolated
  massive stars. This is relevant for understanding the properties of
  zero-metallicity stars (Population III) or stars that form under
  extreme environmental conditions like in the Galactic center or in
  luminous starbursts.
\end{abstract}

\section{Introduction}

Identifying the physical processes that determine the masses of stars
and their statistical distribution, usually known as the initial mass
function (IMF), is a fundamental problem in star formation research.
Observations in the solar vicinity and in nearby young star clusters
suggest the existence of a characteristic stellar mass
$M_{\mathrm{ch}}$.  The IMF peaks at this characteristic mass which is
typically a few tenths of a solar mass. For larger masses the IMF has
a nearly power-law form and it declines rapidly towards smaller masses
(Scalo 1998; Kroupa 2002; Chabrier 2003).  \nocite{SCA98}
\nocite{KRO02} \nocite{CHA03}

Although the IMF in our Milky Way has been derived from vastly
different regions, ranging from the immediate neighborhood of our Sun
to dense and distant stellar clusters, its key features seem to be
strikingly universal to all determinations (e.g.\ Kroupa 2001, Meyer
et al.\ 2004 and references therein). However, the environmental
conditions in the considered regions vary enormously. If the formation
of stars and the resulting mass spectrum strongly depend on initial
conditions, there would thus be no reason for the IMF to be universal.
A derivation of the characteristic stellar mass that is based on
fundamental atomic and molecular physics would therefore be highly
desirable.

%% \nocite{KRO01b} \nocite{LAR03} \nocite{MAC04}
%% \nocite{LAR81} \nocite{FLE82} \nocite{PAD95} \nocite{PAD97}
%% \nocite{KLE98, KLE00b} \nocite{KLE01c} \nocite{PAD02} \nocite{SCH04}
%% \nocite{JAP04} \nocite{BOD95} \nocite{PAP95} \nocite{LIN96}

On the other hand, there are hints for IMF variations in more extreme
environments and/or at different metallicities. For example, HST/WFPC2
observations of the LMC by Gouliermis, Brander \& Henning (2004)
indicate that local conditions seem to favor the formation of higher
mass stars (top-heavy IMF) in associations, and not in the background
field. Similar conclusions may be drawn for the stellar mass spectrum
in the Galactic Center (Mark Morris, this proceedings; or Stolte et
al.\ 2003 for the Arches Cluster) or in starburst galaxies (e.g.\ 
Elmegreen 2004, and references therein).

There are various ways to approach the formation of stars and star
clusters from a theoretical point of view. Currently, the most
successful models are based on the interplay between self-gravity and
the interstellar turbulence ubiquiteously observed in Galactic
molecular clouds. In this picture, stars form by a process we call {\sl
  gravoturbulent fragmentation} (see, e.g.\ the reviews by Larson 2003,
or Mac~Low \& Klessen 2004).  However, despite a wealth of
observational data for the Milky Way (e.g.\ Fuller \& Myers 1992;
Goodman et al.\ 1998), the nature of interstellar turbulence and the
thermodynamic state of the star-forming gas (as characterized by the
equation of state -- EOS) remain major theoretical problems in
understanding the stability and collapse of molecular clouds. The
stiffness of the EOS can be largely responsible for the resulting
density probability function of interstellar gas in the turbulent ISM.
In particular, the value of the polytropic index $\gamma$ when
adopting an EOS of the form $P\propto\rho^\gamma$ strongly influences
the compressibility of density condensations as well as the amount of
clump fragmentation.  Hence, the EOS is directly related to the IMF
(V\'azquez-Semadeni et al.\ 1996; Li, Klessen, \& Mac~Low 2003;
Jappsen et al.\ 2005).  In Spaans \& Silk (2000, 2005) the properties
of a polytropic EOS were investigated and it was found that the
stiffness of the EOS depends strongly on the ambient metallicity and
the infrared background radiation field produced by irradiated dust
grains. It thus may vary considerably in different galactic
environments.

The structure of this paper is as follows: In the next section,
\S\ref{sec:gravoturb-frac}, we briefly review the basic concepts of
gravoturbulent fragmentation and discuss its relation to stellar
birth. In \S\ref{sec:poly-gas} we consider the fragmentation behavior
of self-gravitating, supersonically turbulent gas with a polytropic
EOS. Small values of the polytropic exponent $\gamma$ lead to the
formation of dense clusters of low-mass stars, while $\gamma > 1$
introduces a bias towards more massive and isolated stars. In
\S\ref{sec:EOS-in-Galaxy}, we discuss the thermal properties of the
star-forming gas in our Galaxy and their implications for the IMF in
the solar neighborhood. Finally, in \S\ref{sec:starburst-IMF} we
present first numerical calculations of gravoturbulent fragmentation
in starburst galaxies  suggesting that the IMF in very extreme
environments may be top-heavy.

\section{The Concept of Gravoturbulent Fragmentation}
\label{sec:gravoturb-frac}
Supersonic turbulence plays a dual role in star formation. While it
usually is strong enough to counterbalance gravity on global scales it
will usually provoke collapse locally. This process is discussed in
detail in the reviews by Larson (2003) and Mac~Low \& Klessen (2004).
Turbulence establishes a complex network of interacting shocks, where
regions of high-density build up at the stagnation points of
convergent flows.  These gas clumps can be dense and massive enough to
become gravitationally unstable and collapse when the local Jeans
length becomes smaller than the size of the fluctuation.
%%
%% {\bf comment on micorturbulence}
%% % It should be
%% % mentioned at this place a microturbulent approach (Chandrasekhar 1951)
%% % is not valid in typical molecular clouds
%%
However, the fluctuations in turbulent velocity fields are highly
transient.  They can disperse again once the converging flow fades
away (V{\'a}zquez-Semadeni et al.\ 2005).  Even clumps that are
strongly dominated by gravity may get disrupted by the passage of a
new shock front (Mac~Low et al.\ 1994).

For local collapse to result in the formation of stars, Jeans unstable,
shock-generated, density fluctuations  must collapse to sufficiently
high densities on time scales shorter than the typical time interval between
two successive shock passages.  Only then do they `decouple' from the ambient
flow pattern and survive subsequent shock interactions.  The shorter the time
between shock passages, the less likely these fluctuations are to survive. The
overall efficiency of star formation depends strongly on the wavelength and
strength of the driving source (Klessen et al.\ 2000; Heitsch et al.\
2001). Both regulate the amount of gas available for collapse on
the sonic scale where turbulence turns from supersonic to subsonic
(V\'azquez-Semadeni et al.\ 2003).

The velocity field of long-wavelength turbulence is dominated by
large-scale shocks which are very efficient in sweeping up molecular
cloud material, thus creating massive coherent structures. These
exceed the critical mass for gravitational collapse by far. The
situation is similar to localized turbulent decay, and quickly a
cluster of protostellar cores builds up (e.g.\ Klessen et al.\ 1998;
Klessen 2001; Bate et al.\ 2003; Li et al. 2004; or Bonnell, this
proceedings).  A prominent example is the Trapezium Cluster in Orion
with a few thousand young stars (Hillenbrand 1997).  However, this
scenario also applies to the Taurus star-forming
region (Hartmann 2002) which is historically considered as a case of isolated stellar
birth.  Its stars have formed almost simultaneously within several
coherent filaments which apparently are created by external
compression (see Ballesteros-Paredes et al.\ 1999). This renders it a
clustered star-forming region in the sense of the above definition.

The efficiency of turbulent fragmentation is reduced if the driving
wavelength decreases. There is less mass at the sonic scale and the
network of interacting shocks is very tightly knit. Protostellar cores
form independently of each other at random locations throughout the
cloud and at random times. There are no coherent structures with
multiple Jeans masses. Individual shock generated clumps are of low
mass and the time interval between two shock passages through the same
point in space is small.  Hence, collapsing cores are easily destroyed
again. Altogether star formation is inefficient, and stars are
dispersed throughout the cloud.

\section{Gravoturbulent Fragmentation in Polytropic Gas}
\label{sec:poly-gas}
As indicated above, the thermodynamic properties of the star-forming
gas may play a crucial role in determining the stellar mass spectrum.
As a first and crude approximation, the balance between heating and
cooling in a molecular cloud can be described by a polytropic EOS,
$P=K\rho^{\gamma}$, where $K$ is a constant, and $P$, $\rho$ and
$\gamma$ are thermal pressure, gas density, and polytropic exponent,
respectively. Depending on the environment, the exponent $\gamma$ can cover the
range $0.2 < \gamma < 1.4$ in the interstellar medium (e.g.\ Spaans \&
Silk 2000, 2005; Scalo \& Biswas 2002).

\begin{figure}[t]
 \centerline{\epsfxsize=0.32\textwidth\epsffile{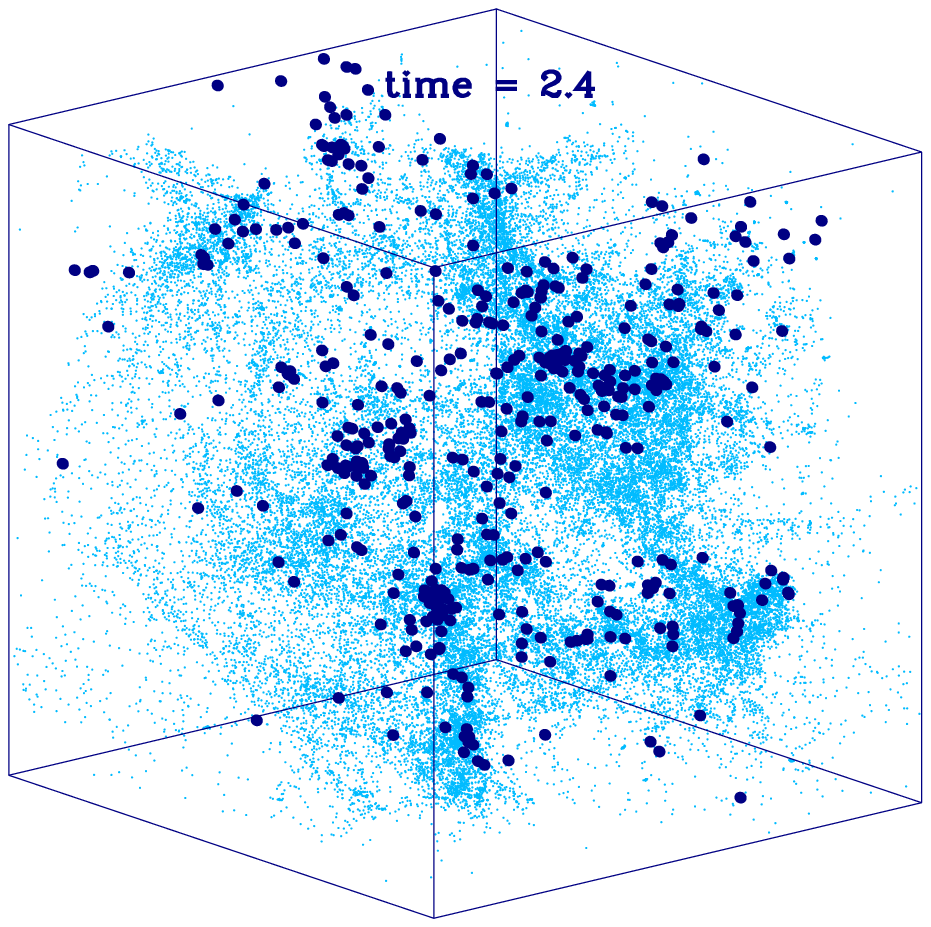}
             \epsfxsize=0.32\textwidth\epsffile{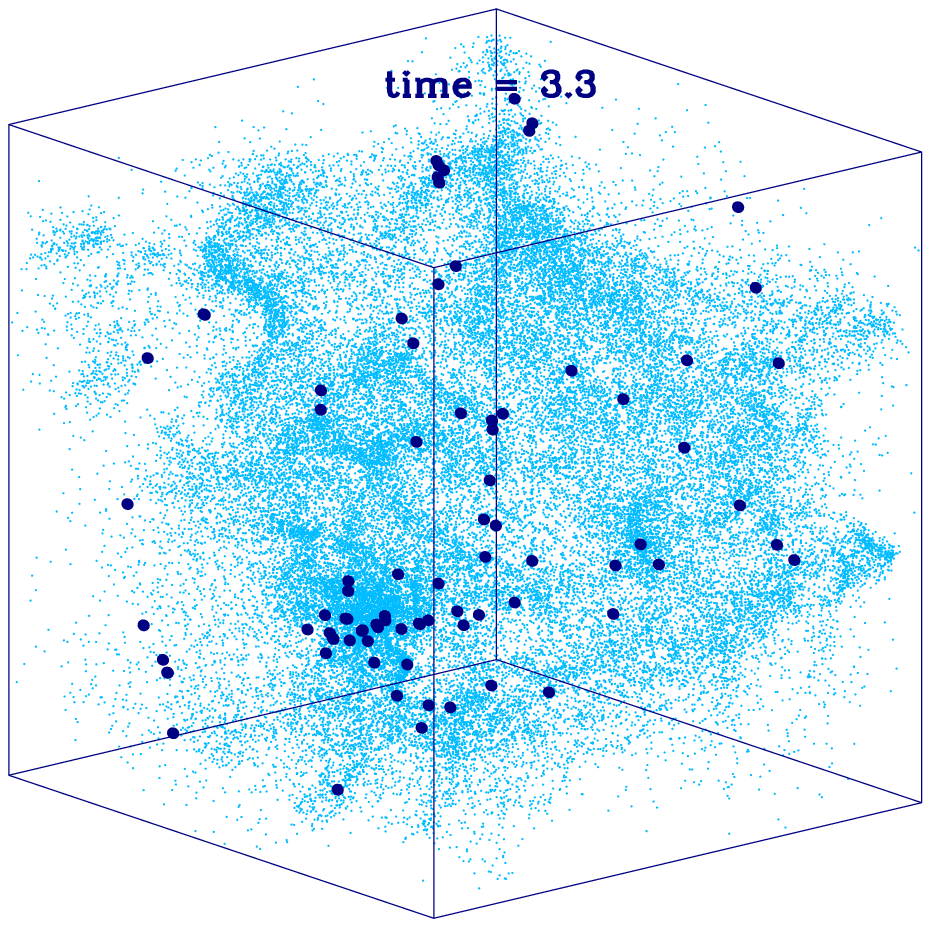}
             \epsfxsize=0.32\textwidth\epsffile{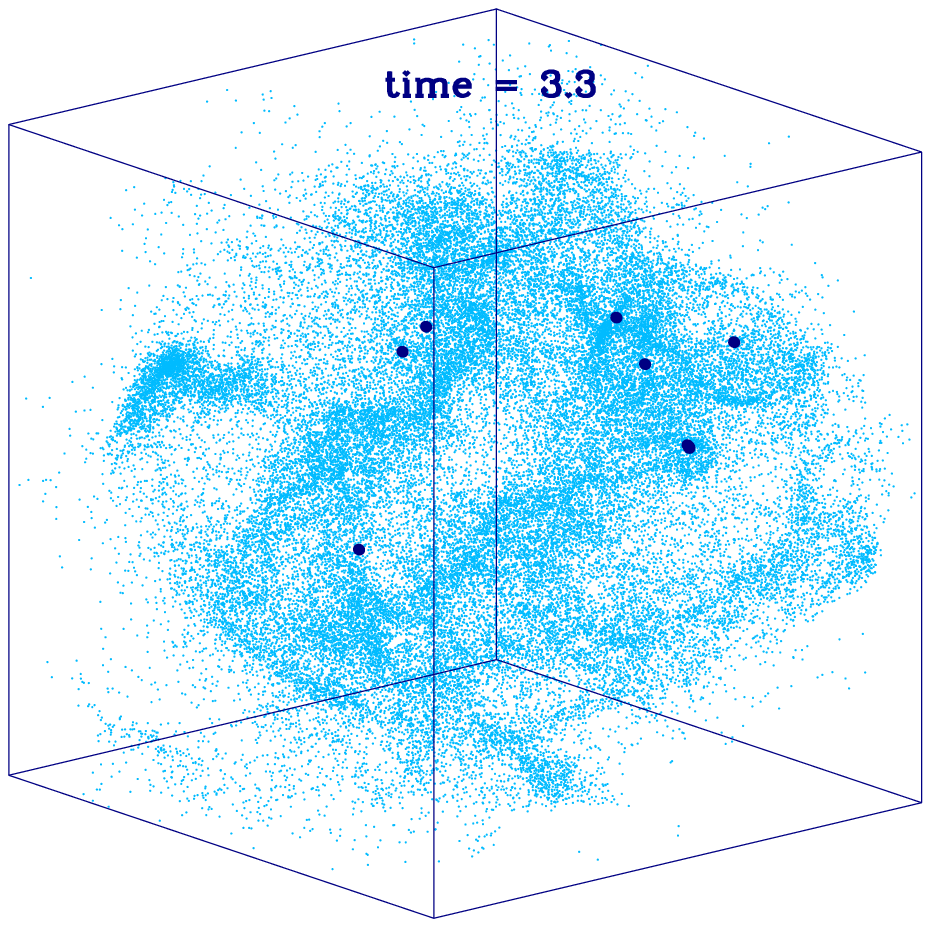}}
 \centerline{\epsfxsize=0.32\textwidth\epsffile{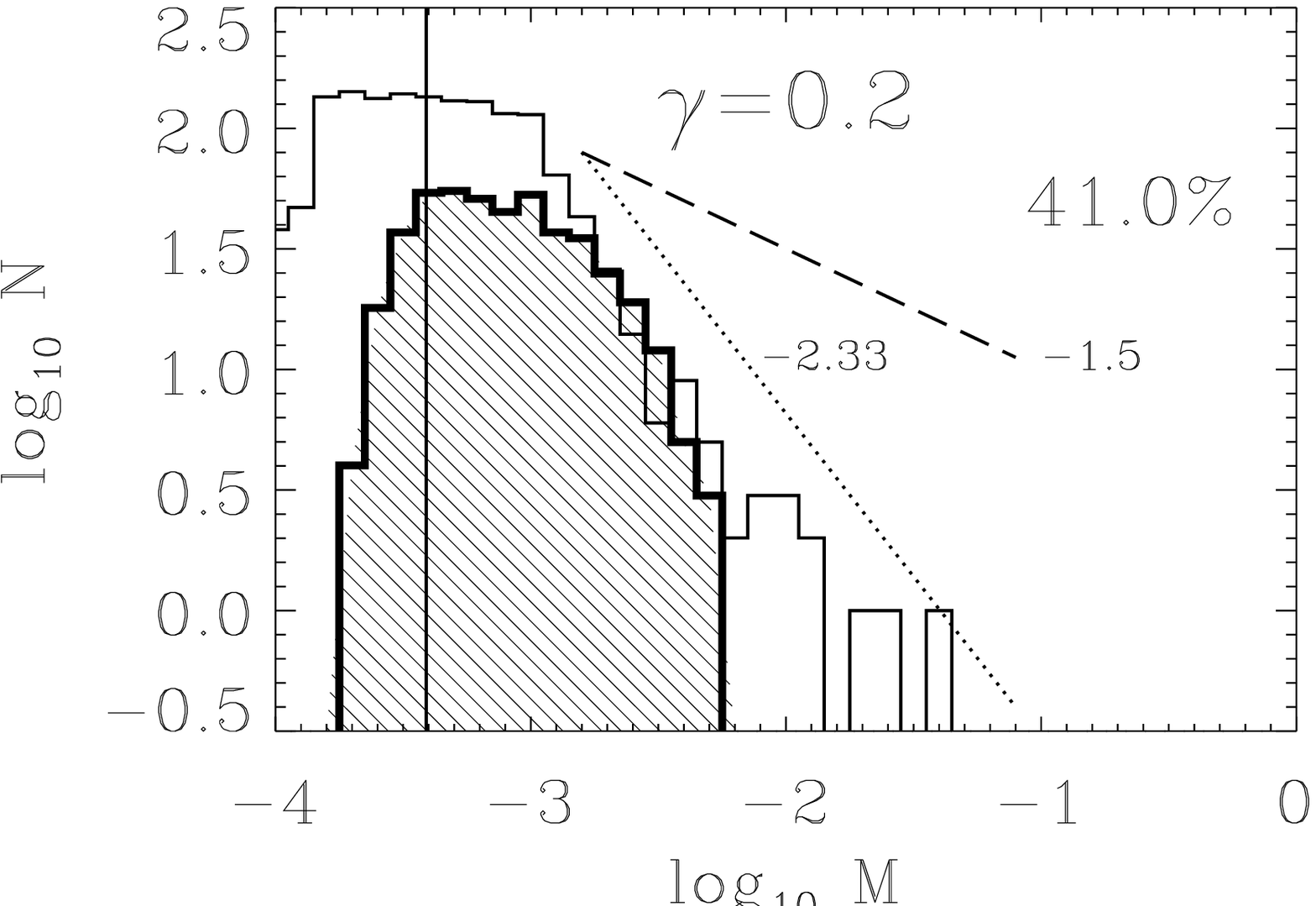}
             \epsfxsize=0.32\textwidth\epsffile{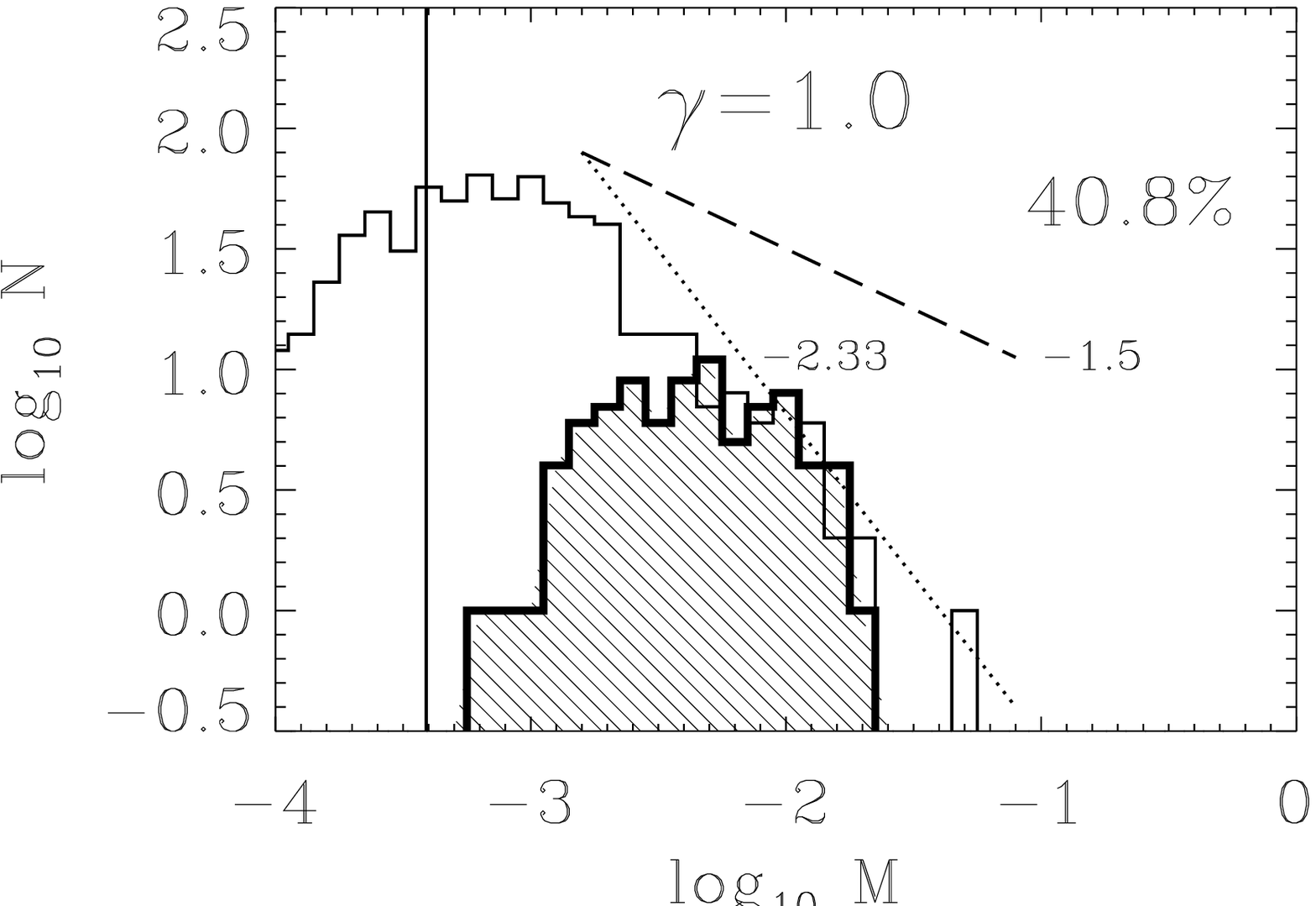}
             \epsfxsize=0.32\textwidth\epsffile{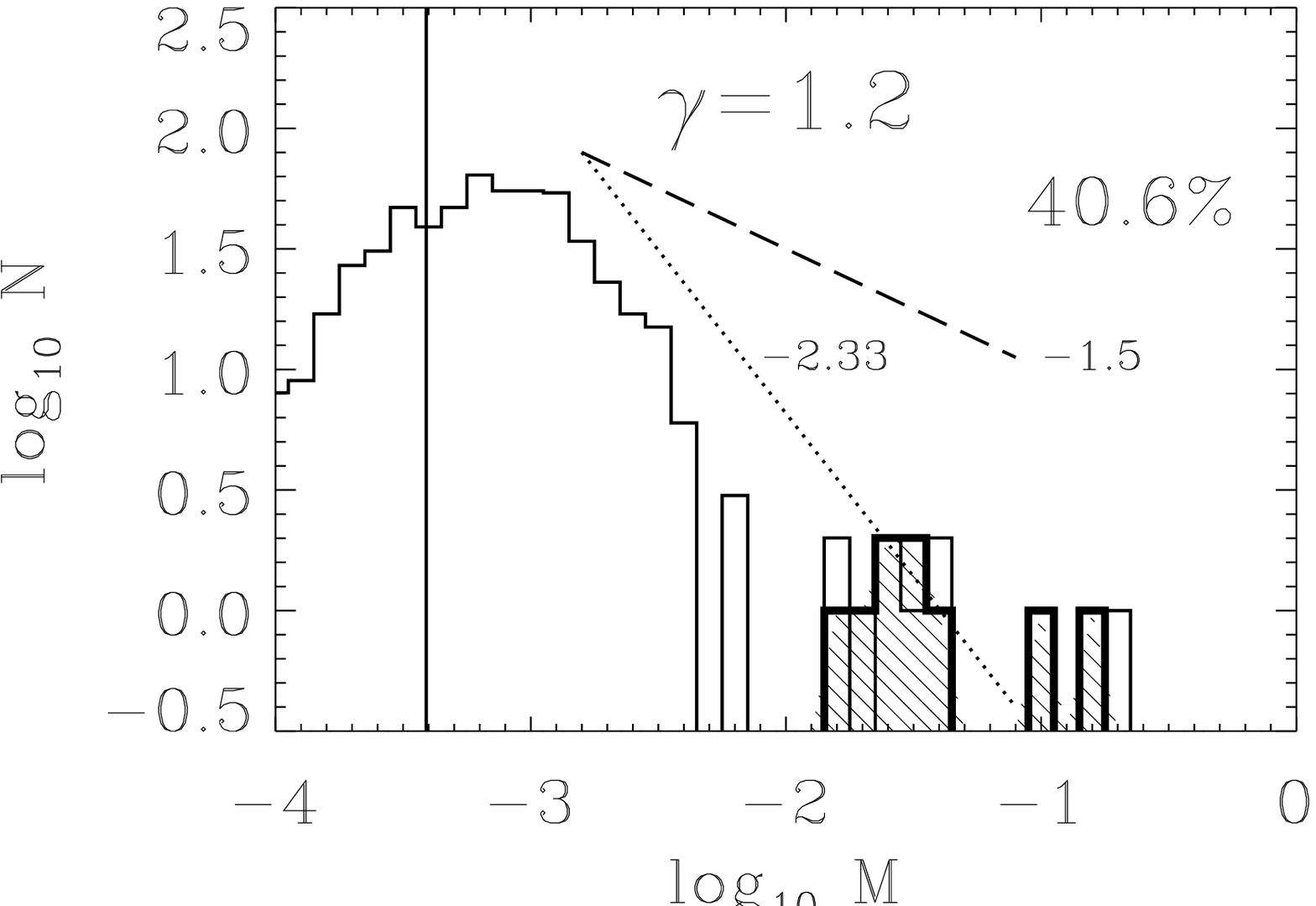}}
\caption{Top: 3-D distribution of the gas and protostars for
    different $\gamma$. Bottom: Mass spectra of gas clumps
    (\textit{thin  
    lines}) and of protostars (collapsed cores: \textit{hatched thick-lined
    histograms}) for the corresponding cube above. The
    percentage shows the fraction of total mass accreted onto protostars. The
    vertical line shows the numerical resolution limit. Shown also are two
    power-law spectra with $\nu = -1.5$ (dashed-line) and $\nu =
    -2.33$ (dotted line). Figure adopted from \cite{LI03}.}
\label{fig:fig-gamma}
\end{figure}

\cite{LI03} carried out detailed numerical calculations to determine
the effects of different EOS's on gravoturbulent fragmentation by
varying $\gamma$ in steps of 0.1 in otherwise identical simulations.
Figure \ref{fig:fig-gamma} illustrates how a low exponent $\gamma$
leads to the build-up of a dense cluster of stars, while high values
of $\gamma$ result in isolated star formation. It also shows that the
spectra of both the gas clumps and protostars change with $\gamma$. In
low-$\gamma$ models, the mass distribution of the collapsed
protostellar cores at the high-mass end is roughly log-normal. As
$\gamma$ increases, fewer but more massive cores emerge. When $\gamma
> 1.0$, the distribution is dominated by high mass protostars only,
and the spectrum tends to flatten out. It is no longer described by
either a log-normal or a power-law. The clump mass spectra, on the
other hand, do show power-law behavior at the high mass side, even for
$\gamma > 1.0$.

This suggests that in a low-$\gamma$ environment stars tend to form in
clusters and with small masses. On the other hand, massive stars can
form in small groups or in isolation in gas with $\gamma > 1.0$. 

The usefulness of theoretical predictions for the formation of
isolated massive stars may be debated, since usually, massive stars are
found in clusters. It is therefore interesting to note, that recently
\citet{lamers02} reported observations of isolated massive stars or
very small groups of massive stars in the bulge of M51. Also
\cite{massey02} finds massive, apparently isolated field stars in both
the Large and Small Magellanic Clouds. This is consistent with our
models assuming $\gamma > 1.0$.
 
%% High resolution simulations by \cite{kitsionas_abel02} of the
%% formation of Population III stars suggest that in very metal-deficient
%% gas only one massive object forms per pregalactic halo. In the early
%% Universe, inefficient cooling due to the lack of metals may result in
%% high $\gamma$. Our models then suggest weak fragmentation, supporting
%% the hypothesis that the very first stars build up in isolation.

%% Finally, a recent investigation by \cite{kitsionas_jappsen05} has
%% shown that the position of the peak of the core mass spectrum does not
%% only depend on the {\em average thermal Jeans mass} $\langle M_{\rm
%%  J}\rangle$ of pre-stellar gas but also on a characteristic mass defined
%% by the density at which the polytropic index $\gamma$ changes value 
%% from smaller to larger than one.

\section{Thermal Properties of Galactic Gas and
  Implications for the IMF}
\label{sec:EOS-in-Galaxy}
Observational evidence indicates that dense prestellar cloud cores show
a rough balance between gravity and thermal pressure \citep{BEN89,
  MYE91}.  Thus, the thermodynamic properties of the gas play an
important role in determining how dense star-forming regions in
molecular clouds collapse and fragment.  Observational and theoretical
studies of the thermal properties of collapsing clouds both indicate
that at densities below $10^{-18}\,\mathrm{g\,cm^{-3}}$, roughly
corresponding to a number density of
$n=2.5\times10^5\,\mathrm{cm^{-3}}$, the temperature decreases with
increasing density. This is due to the strong dependence of molecular
cooling rates on density \citep{KOY00}. Therefore, the polytropic
exponent $\gamma$ is below unity in this density regime. At densities
above $10^{-18}\,\mathrm{g\,cm^{-3}}$, the gas becomes thermally
coupled to the dust grains, which then control the temperature by
far-infrared thermal emission. The balance between compressional
heating and thermal cooling by dust causes the temperature to increase
again slowly with increasing density. Thus the temperature-density
relation can be approximated with $\gamma$ above unity in this regime
\citep{LAR85,LAR05,SPA00,SPA05}. Changing $\gamma$
from a value below unity to a value above unity results in a minimum
temperature at the critical density. As shown by \citet{LI03}, gas
fragments efficiently for $\gamma < 1.0$ and less efficiently for
higher $\gamma$. Thus, the Jeans mass at the critical density defines
a characteristic mass for fragmentation, which may be related to the
peak of the IMF.

\begin{figure}[t!]          
\centerline{\epsfxsize=0.75\textwidth\epsffile{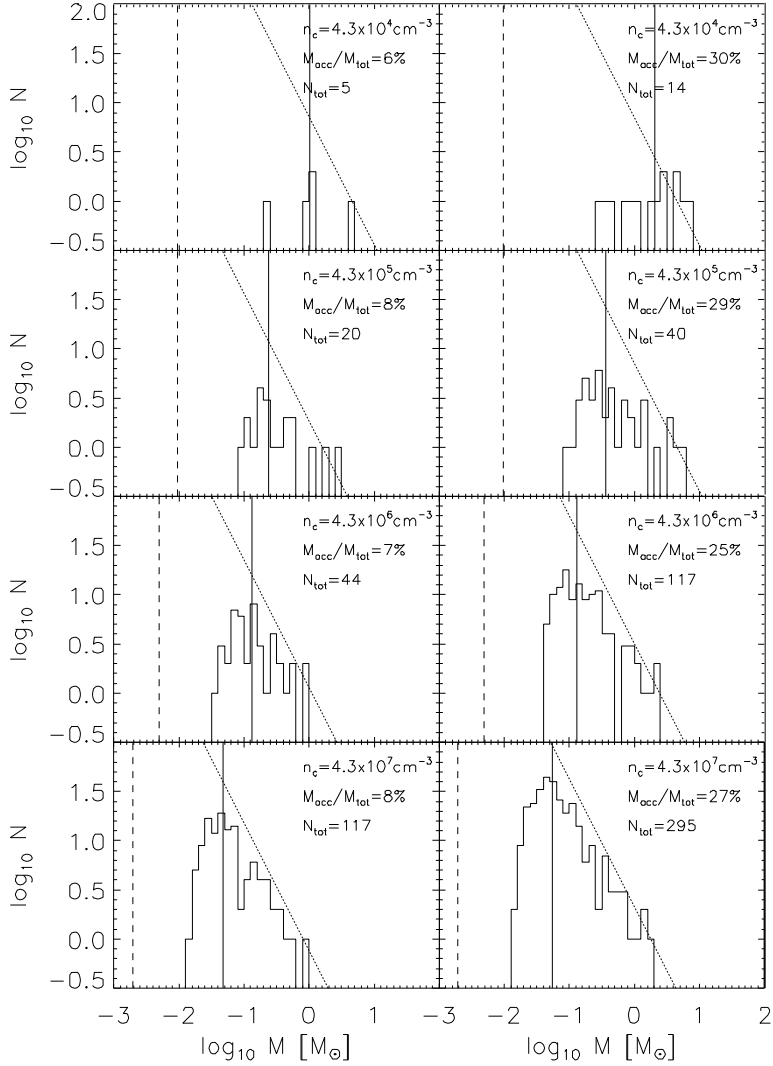}}
\caption{Mass spectra of protostellar cores for four models with critical
 densities in the range $4.3\times10^4\,\mathrm{cm^{-3}} \le n_{\mathrm{c}} \le
4.3\times10^7\,\mathrm{cm^{-3}}$. We show two phases of evolution, when about $10\%$ and
  $30\%$ of the total mass has been accreted onto protostars. The {\it vertical solid line \nocorr}shows the position of
  the median mass. The {\it dotted line \nocorr}serves as a reference to the Salpeter value \citep{SAL55}. The {\it dashed
  line \nocorr}indicates the mass resolution limit. (Adopted from Jappsen et
al.\ 2005.)}
\label{fig:all-mspec}
\end{figure}

This hypothesis has been tested by \cite{JKLLM05}, who studied the
effects of using a piecewise polytropic EOS with a polytropic exponent
that changes from $\gamma = 0.7$ to $\gamma=1.1$ at critical densities
in the range $4.3\times10^4\,\mathrm{cm^{-3}} \le n_{\mathrm{c}} \le
4.3\times10^7\,\mathrm{cm^{-3}}$. Figure \ref{fig:all-mspec}
illustrates how the characteristic mass 
 $M_{\mathrm{ch}}$
 for fragmentation depends on
the adopted critical density. Closest correspondence with the observed
IMF \citep[see, ][]{SCA98, KRO02, CHA03} occurs for a critical density of
$4.3\times10^6\,\mathrm{cm^{-3}}$ and for stages of accretion around
$30\%$ and above. At high masses, the distribution of collapsed objects follows a
Salpeter-like power law. For comparison we indicate the Salpeter slope
$x\approx1.3$ \citep{SAL55} where the IMF is defined by
$\mathrm{d}N/\mathrm{d}\log m \propto m^{-x}$. For masses around the
median mass the distribution exhibits a small plateau and then falls
off towards smaller masses.
The top-row model where the change in $\gamma$ occurs below the
initial mean density, shows a flat distribution with only a few, but
massive cores. They reach masses up to $10\,$M$_{\odot}$ and the minimal
mass is about $0.3\,$M$_{\odot}$.  The distribution becomes more peaked
for higher $n_{\mathrm{c}}$ and there is a shift to lower masses.

Altogether, the characteristic mass~$M_{\mathrm{ch}}$ decreases with
increasing critical density~$n_{\mathrm{c}}$ following the relation
$M_{\mathrm{ch}} \propto n_{\mathrm{c}}^{-0.5\pm0.1}$.  The density at
which $\gamma$ changes from below unity to above unity thus selects a
characteristic mass scale for fragmentation. For physical parameters
appropriate for the local interstellar medium, the resulting mass
spectrum exhibits striking similarities to the observed IMF.

\section{The IMF in Starburst Galaxies}
\label{sec:starburst-IMF}
In this section, we present a preliminary and first approach to
understanding the IMF in starburst galaxies using direct numerical
simulations of gravoturbulent cloud fragmentation.

Again, we employ a polytropic EOS, $P\propto\rho^{\gamma}$, where
$\gamma$ is the polytropic index and $\rho$ the mass density.  We
adopt a perfect gas equation of state, $P\propto\rho T_g$ for the gas
temperature $T_g$, to write $\gamma$ as
$$\gamma =1+{{d{\rm log}T_g}\over{d{\rm log}\rho}}.$$
This last step
is justified (Scalo \& Biswas 2002; V\'azquez-Semadeni, Passot \&
Pouquet, 1996) as long as the heating and cooling terms in the fluid
energy equation can adjust to balance each other on a time-scale
shorter than the time-scale of the gas dynamics (i.e., thermal
equilibrium).  The model described in Spaans \& Silk (2000, 2005) is
used in this work and the interested reader is referred to these
papers for a detailed description of the various heating and cooling
terms that influence the polytropic index.

The EOS was computed for self-gravitating  spherical clouds
approximated as singular isothermal spheres (Neufeld, Lepp \& Melnick 1995),
for which the total hydrogen number density
$n_{\rm H}$ scales with column density $N_{\rm H}$ per unit of velocity as
$$N_{\rm H}=7.2\times 10^{19}n_{\rm H}^{0.5} {\rm cm}^{-2}/({\rm km\,
  s}^{-1}).$$
For the starburst environments of interest here, we
assume a cosmic ray ionization rate of $3\times 10^{-15}$ s$^{-1}$.
Solar relative abundances are assumed (Asplund et al., 2004; Jenkins
2004) while the overall metallicity is two times solar (Barthel 2004).
We further adopt a MRN grain size distribution (Mathis, Rumpl \&
Nordsieck 1977) and assume that the dust abundance scales with the
carbon abundance.  For the starburst model $\Delta V_{\rm tur}=3$ km/s
is adopted to take the larger input of kinetic energy (e.g.\ through
supernovae) into account. The dust temperature inside the model clouds
is set by a fiducial background star formation rate of 100 $M_\odot$
yr$^{-1}$ which causes dust to be at temperatures of about $T_d=70$ to
$T_d=30$ K, depending on the amount of shielding. These values are
consistent with dust temperatures determined by continuum observations
of luminous infrared galaxies (Spinoglio et al., 2002; Klaas et al.\ 
2001, 1997).  No freeze-out of molecules is assumed. The latter
assumption is justified because our dust is generally warmer than 20 K.

\begin{figure}[t]          
\centerline{\epsfxsize=0.70\textwidth\epsffile{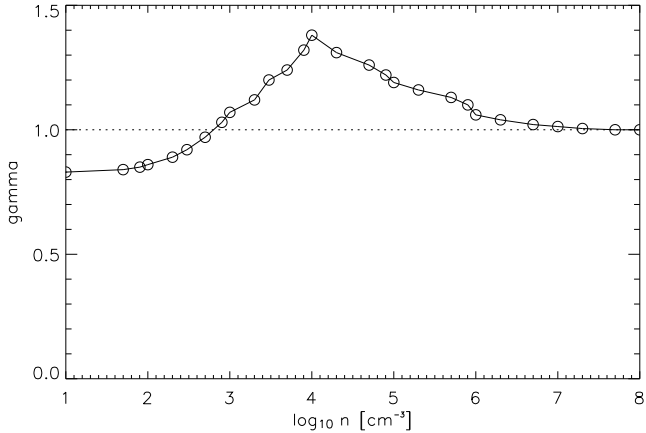}}
\caption{Starburst EOS adopted from Spaans \& Silk (2005).}
\label{fig:starburst-EOS}
\end{figure}
An overabundance, relative to solar, of oxygen leads to a
significantly stiffer EOS because the opacity in oxygen bearing
molecules like CO and water is higher. This increases the amount of
line trapping in these opaque cooling lines. Furthermore, when the
temperature of the dust is high (larger than 30 K), H$_2$O collisional
de-excitation heating dominates the thermal budget because the water
levels are pumped by far-infrared dust emission. This causes the gas
heating to attain a $n_{\rm H}^2$ density dependence (Takahashi,
Hollenbach \& Silk 1983; Spaans \& Silk 2000, 2005).  Consequently,
one has a strong peak, $\gamma\approx 1.4$, around a density of $\sim
10^4$ cm$^{-3}$.  Since the dust grains are warm, gas-dust heating
around $10^5$ cm$^{-3}$ causes the EOS to remain above unity,
$\gamma\approx 1.1$, beyond the H$_2$O opacity peak. Note here that
gas-dust heating/cooling scales as $(T_g-T_d)T_g^{0.5}n_{\rm H}^2$.
Hence, a starburst has a stiff EOS because of its high absolute
metallicity and warm, $T_d\sim 30-70$ K, dust.  Figure
\ref{fig:starburst-EOS} shows the resulting EOS and the fit that has
been adopted for the hydrodynamic simulation.

We adopt this EOS to model gravoturbulent star formation in a
circumnuclear starburst environment using smoothed particle
hydrodynamics (SPH).  This is a Lagrangian method to solve the
equations of hydrodynamics, where the fluid is represented by an
ensemble of particles, and flow quantities are obtained by averaging
over an appropriate subset of SPH particles (Benz 1990; Monaghan
1992).  We use the parallel code GADGET, designed by \citet{SPR01}.
The method is able to resolve large density contrasts as particles are
free to move, and so the particle concentration increases naturally in
high-density regions. To determine the resolution limit of our
calculation we deploy a criterium proposed by \citet{BAT97}. It
requires the local Jeans mass at all densities to be represented by at
least 100 particles, and is adequate for the problem considered here,
where we follow the evolution of highly nonlinear density fluctuations
created by supersonic turbulence. Our version of the code furthermore
replaces high-density cores with sink particles \citep*{BAT95}. Sink
particles can accrete gas from their surroundings while keeping track of
mass and momentum. This enables us to follow the dynamic evolution of
the system over many local free-fall timescales.

\begin{figure}[t]          
\centerline{\epsfxsize=0.6\textwidth\epsffile{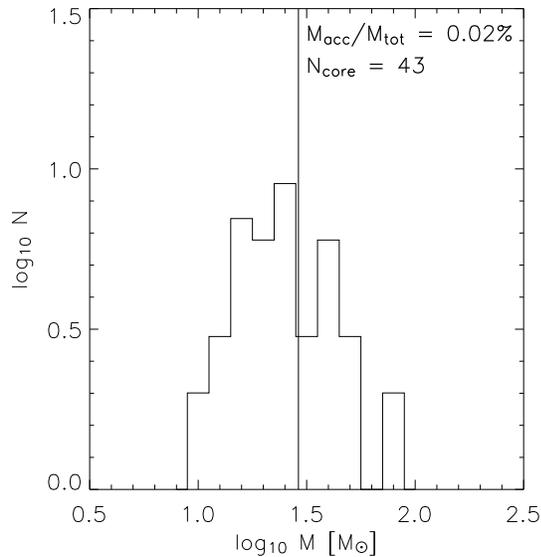}}
\caption{Mass spectrum of protostellar objects in a simulation
  focusing on a starburst environment. The size of the considered
  molecular cloud region is $11.2\,$pc and it contains
  $80\,000\,$M$_{\odot}$. The system is depicted at early times when
  about 0.02\% of the available gas is converted into stars.
  Comparison with Fig.\ \ref{fig:all-mspec} shows that the average
  mass (indicated by the vertical line) is shifted towards high
  values. The nominal resolution limit lies at $1\,$M$_{\odot}$.
}
\label{fig:starburst-IMF}
\end{figure}
We focus on a cubic molecular cloud region of $11.2\,$pc in size. The
region contains $80\,000\,$M$_{\odot}$ of gas and has a mean particle
density $n=1\,000\ $cm$^{-3}$. We drive turbulence uniformly on large
scales, with wave numbers $k$ in the range $1 \le k \le 2$ as
described in Mac~Low (1999).  The energy input rate is adjusted to
yield a constant turbulent Mach number
$\mathcal{M}_{\mathrm{rms}}\approx 5$. The particle number in the
simulation presented here is $N=8\,000\,000$, which is thus one of the
highest resolution star-formation calculation ever done with SPH. The
critical density for sink particle formation is $n_c =
10^7\,$cm$^{-3}$. The mass of individual SPH particles is
$m=0.01\,$M$_{\odot}$ which is sufficient to resolve the minimum Jeans
mass in the system $M_{\rm J} \approx 1.5\,$M$_{\odot}$. Except for
the EOS and the particle number, the numerical set-up is identical to
the study by \cite{JKLLM05}. 

As a preliminary result, we find that in the considered starburst
galaxy, the mass spectrum of protostellar objects is biased towards
high masses. This is illustrated in Fig.\ \ref{fig:starburst-IMF}.
The calculation thus supports the hypothesis that for extreme
environmental conditions such as inferred for very dusty and
IR-luminous circumnuclear starbursts the stellar initial mass function
may indeed  be top-heavy.

\begin{acknowledgments}
  We thank Javier Ballesteros-Paredes, Peter Bodenheimer, Spyros
  Kitsionas, Richard Larson, Yuexing Li, Mordecai Mac~Low, Stefan
  Schmeja, Joe Silk, and Enrique V\'{a}zquez-Semadeni for stimulating
  discussions and collaborations. RSK acknowledges support by the Emmy
  Noether Program of the Deutsche Forschungsgemeinschaft (grant no.\ 
  KL1358/1).
\end{acknowledgments}

\end{document}